\documentclass[prb,aps,epsf,twocolumn,showpacs,superscriptaddress]{revtex4}
\usepackage[dvips]{graphicx}
\usepackage{latexsym}
\usepackage{amsmath}
\begin{document}

\title{Theory of a continuous Mott transition in two dimensions}

\author{T. Senthil}
\affiliation{ Department of Physics, Massachusetts Institute of
Technology, Cambridge, Massachusetts 02139}

\date{\today}
\begin{abstract}
We study theoretically the zero temperature phase transition in two dimensions from a Fermi liquid to a paramagnetic Mott insulator with a spinon Fermi surface. We show that the approach to the bandwidth controlled Mott transition from the metallic side is accompanied by a vanishing quasiparticle residue and a diverging effective mass. The Landau parameters $F^0_s, F^0_a$ also diverge. Right at the quantum critical point there is a sharply defined `critical Fermi surface' but no Landau quasiparticle. The critical point has a $Tln\frac{1}{T}$ specific heat and a non-zero $T = 0$ resistivity.
We predict an interesting {\em universal resistivity jump} in the residual resistivity at the critical point as the transition is approached from the metallic side. The crossovers out of the critical region are also studied. Remarkably the initial crossover out of criticality on the metallic side is to a Marginal Fermi Liquid metal. At much lower temperatures there is a further crossover into the Landau Fermi liquid. The ratio of the two crossover scales vanishes on approaching the critical point. Similar phenomena are found in the insulating side. The filling controlled Mott transition is also studied. Implications for experiments on the layered triangular lattice organic material $\kappa-(ET)_2Cu_2(CN)_3$ are discussed.

\end{abstract}
\newcommand{\be}{\begin{equation}}
\newcommand{\ee}{\end{equation}}
\maketitle
\section{Introduction}
Despite several decades of work the Mott transition between a metal and an insulator in two or three dimensional systems remains poorly understood\cite{mttrev}. The nature of the transition quite possibly depends on the structure of the insulating state that obtains in the vicinity of the metal-insulator transition. The most familiar example of a Mott insulator is one that has magnetic long range order but it has long been known that other forms of order (like spin-Peierls) are possible as well. An exciting possibility is a Mott insulator with no conventional order that does not break any symmetries\cite{pwarvb}. Such states (popularly called quantum spin liquids) are now known to be theoretically possible\cite{sltheory,rantner,stableu1}. Recent experiments find encouraging evidence for the existence of spin liquid states in a few different systems\cite{kanoda0,yslee,takagi}.

In a number of materials the Mott transition is  (at $T = 0$) first order\cite{mttrev}. In several systems however strong fluctuation effects attributable to the impending localization are seen in metals that are close to the Mott transition\cite{mttrev}. An important and interesting question is whether second order Mott transitions are at all possible. Recent
experiments on the two dimensional organic material $\kappa-(ET)_2Cu_2(CN)_3$ provide tantalizing hints that such a second order Mott transition may actually be possible. At ambient pressure this is a Mott insulator which does not have magnetic long range order down to the lowest temperatures (much smaller than the spin exchange energy)\cite{kanoda0}. Remarkably the spin susceptibility is a non-zero constant at low temperature and the specific heat is linear exactly as in a metal\cite{etspht}. Thus the Mott insulator retains some of the characteristics of a fermi liquid metal. These results have been interpreted\cite{lesik,leesq} as evidence for a quantum spin liquid with gapless neutral fermionic spin-$1/2$ spinon excitations. Theoretically such spin liquid behavior may be expected if there is considerable amount of virtual charge fluctuations. This may be roughly modeled as a large multiple ring exchange term in a spin model. Such ring exchange terms are known to promote spin liquid behavior\cite{lesik}. In experiments optical transport shows that there is substantial conduction down to reasonably low frequency somewhat below the charge gap\cite{kanoda2}. Under pressure a Mott transition to a metal is observed\cite{kanoda1}. All of these suggest that the ambient pressure insulator is close to being a metal. Thus perhaps the Mott transition in this material is a continuous one.

\begin{figure}
\includegraphics[width=8cm]{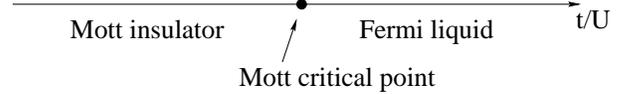}
\caption{Possible schematic zero temperature phase diagram for a half-filled single band repulsive Hubbard model on a non-bipartite lattice. $U$ is the Hubbard interaction strength and $t$ is the hopping amplitude. } \label{mottpdia}
\end{figure}

Theoretically  $\kappa-(ET)_2Cu_2(CN)_3$ is believed to be described by a one band Hubbard model on an isotropic triangular lattice. In an important development the Mott transition at fixed filling on such lattices was studied by Florens and Georges\cite{florens} within an approximate slave particle mean field treatment of the Hubbard model. In this mean field calculation the Mott transition is second order. As it is approached from the Fermi liquid side the quasiparticle residue vanishes but the effective mass stays constant.

\begin{figure}
\includegraphics[width=8cm]{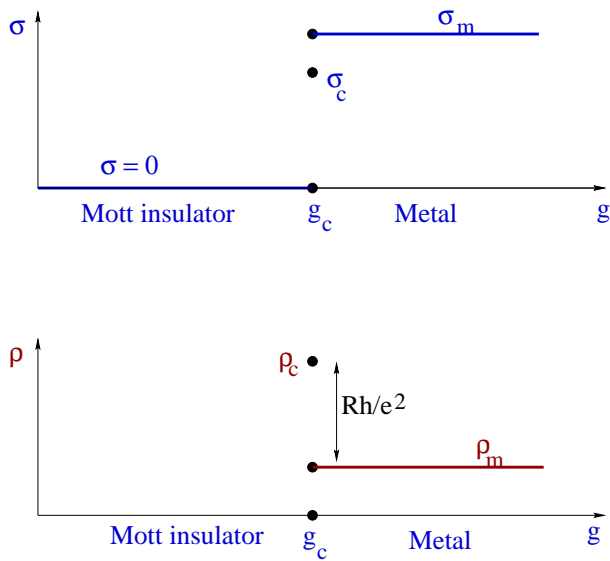}
\caption{Evolution of the extrapolated $T = 0$ conductivity across the Mott transition. The conductivity $\sigma$ jumps twice - once from its value $\sigma_m$ in the metal to its value $\sigma_c$ at the critical point, and then again to zero on moving to the insulating phase. Neither of the jumps are universal. However the jump in the in-plane sheet {\em resistivity} $\rho$ on going from the metal to the Mott critical point is a universal constant $\frac{Rh}{e^2}$ with $R$ of order $1$.}
 \label{Rjmp}
\end{figure}

In this paper we will go beyond this mean field calculation by including fluctuations. Deep in the insulating side fluctuation effects are well understood, and  a gapless spin liquid with a spinon fermi surface and associated gauge fields obtains. Indeed this state has been proposed\cite{lesik,leesq} to describe the Mott insulating phase of $\kappa-(ET)_2Cu_2(CN)_3$. Here we consider the vicinity of the Mott transition itself with a focus on the approach from the metal. We show that
the quasiparticle effective mass of the metal diverges at this Mott transition once fluctuation effects are included. Despite this diverging mass the electronic compressibility vanishes, and the spin susceptibility stays constant. This signals the divergence of the Landau parameters $F^0_{s,a}$. We calculate various physical quantities right at the quantum critical point associated with this Mott transition. In particular the specific heat is shown to behave as $Tln(1/T)$. The resistivity is shown to be a constant which is non-universal. However as the transition is approached from either the metal or insulator the residual resistivity jumps. On approaching from the metal this {\em resistivity jump} is a universal number of order $h/e^2$. (See Fig. \ref{Rjmp}). This prediction can possibly be checked in future experiments.

In a recent paper\cite{critfs} we argued that at a continuous Mott transition the Fermi surface will remain sharply defined even though the Landau quasiparticle does not survive. We demonstrated this within the mean field treatment of Ref. \onlinecite{florens} by a direct calculation of the electron spectral function. Here we include fluctuation effects and show that the sharp critical Fermi surface continues to exist. Further we explicitly show that the electron spectral function $A_c(\vec K, \omega)$ at criticality obeys the scaling ansatz  of Ref. \onlinecite{critfs} with the exponents $z = 1^+$ and $\alpha = -\eta$ (where $\eta$ is the anamolous exponent of the boson field at the $3D$ XY fixed point). The exponent value $1^+$ means that expressions like $\omega^{\frac{1}{z}}$ should be replaced by $\omega ln\frac{1}{\omega}$. Specifically we show that
\begin{equation}
A_c(\vec K, \omega) \sim \frac{\omega^\eta}{ln\frac{\Lambda}{\omega}}F\left(\frac{\omega ln\frac{\Lambda}{\omega}}{v_{F0}k_{\|}}\right)
\end{equation}
right at the Mott transition. Here $k_{\|}$ is the deviation of $\vec K$ from the Fermi momentum along a direction parallel to the normal to the critical Fermi surface at the point of closest approach. Thus the Mott transition studied in this paper provides a concrete example of  a critical Fermi surface.

\begin{figure}
\includegraphics[width=8cm]{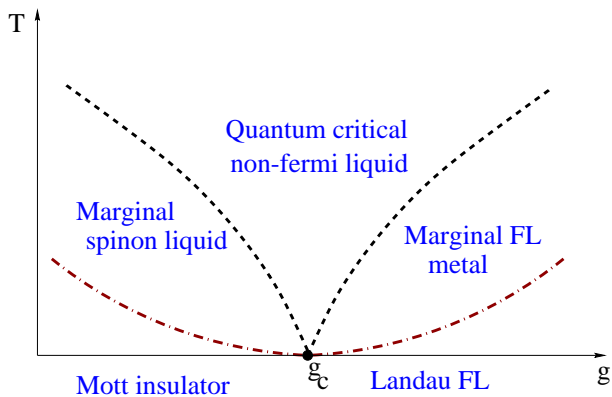}
\caption{Schematic phase diagram showing finite temperature crossovers near the Mott transition studied in this paper.  The black dotted lines represent the crossover scale $T^*$ and the red dotted lines the crossover at $T^{**}$. The quantum critical metal at $T = 0$ has a sharp critical Fermi surface. The electron spectral function at the Fermi surface sharpens into the marginal Fermi liquid form on cooling through $T^*$. It eventually acquires the usual Landau quasiparticle peak only below the much lower scale $T^{**}$. The Mott insulating ground state is a spin liquid with a spinon Fermi surface. A different spin liquid which also has a spinon Fermi surface appears in the intermediate temperature regime in the insulator.  The critical Fermi surface evolves into the spinon Fermi surface in the insulator.} \label{mtttrsn}
\end{figure}

We also study the crossover out of the critical region into either the Fermi liquid or the spin liquid Mott insulator. Remarkably we find that on the metallic side the initial crossover is not to a Landau Fermi liquid but rather to a marginal Fermi liquid metal\cite{varma} (see Fig. \ref{mtttrsn}). As the  tuning parameter $g$ moves away from its critical value $g_c$ into the metallic side,  this happens at an energy scale
\begin{equation}
T^* \sim |g - g_c|^{\nu}
\end{equation}
with $\nu \approx 0.67$ a universal exponent (equal in fact to the correlation exponent of the $3D$ XY model).
The marginal Fermi liquid metal crosses over to a Landau Fermi liquid at a much lower energy scale
\begin{equation}
T^{**} \sim |g - g_c|^{2\nu}
\end{equation}
The ratio $T^{**}/T^*$ thus vanishes on approaching the critical point.
Similar phenomena happen in the insulating side as well. The initial crossover (at energy scale $T^*$) out of the
quantum critical point is to a marginal spinon liquid insulator. In this insulator the spinons have a sharp fermi surface and a scattering rate proportional to the energy. The specific heat is $Tln\frac{1}{T}$. This eventually crosses over to the spinon non-fermi liquid state of Refs. \onlinecite{lesik,leesq} at the much lower scale $T^{**}$.

The presence of these two scales means that a universal scaling function for the initial crossover out of criticality  will only  describe the marginal Fermi liquid state. The low energy physics of the Fermi liquid is not part of this scaling and is obtained only on including the second crossover at $T^{**}$. Thus the scaling hypotheses of Ref. \onlinecite{critfs} is not expected to directly describe the approach to criticality at this particular transition. We show that this is indeed the case.
Similar phenomena are well known in other simpler classical and quantum phase transitions with ``dangerously irrelevant" perturbations. In such cases as in the present problem the usual scaling only describes the initial crossover, and needs to be modified to handle the second one.

Our theory for this Mott transition is formulated in terms of a charge-$0$ spin-$1/2$ fermionic spinon field $f_{\alpha}$, a charge-$e$ spin-$0$ boson field $b$, and an associated $U(1)$ gauge field. We argue that at low energies near the critical point the boson field is dynamically decoupled from the spinon-gauge system. Further the critical properties of this boson are controlled by the usual $3D$ XY fixed point. The spinons and the gauge fields on the other hand form a strongly coupled system. Remarkably in the presence of the critical boson field the mathematical structure of this strongly coupled system is identical to that arising in the Halperin-Lee-Read composite fermi liquid state\cite{hlr} that describes the half-filled Landau level. Specifically our problem maps on the case with $\frac{1}{r}$ Coulomb interactions studied in Ref. \onlinecite{hlr}. We can thus make use of the existing understanding of this state to calculate critical properties of the Mott transition in detail.

We can now explain the origin of the two scales $T^*$ and $T^{**}$ that appear on moving away from criticality. The former is when the boson field crosses over from criticality into either a condensed or insulating phase. However the decoupling between the bosons and the spinon-gauge system continues beyond this scale. It is only at the lower scale $T^{**}$ that the change in the boson dynamics from the critical point is felt by the spinon-gauge system. Loosely speaking we may say that the coupling between the bosons and the spinon-gauge system plays the role of a dangerously irrelevant perturbation at the critical fixed point. However at present we do not have a suitable renormalization group formulation to make this more precise.

We also briefly study the filling controlled Mott transition ({\em i.e} tuned by a chemical potential) to the same spin liquid Mott insulator. Several properties of this transition have already been studied in prior work on the related Kondo breakdown model\cite{svsrev}. We
show that both $Z$ and the inverse effective mass vanish as the transition is approached from the metal. The Landau parameters $F^0_{s,a}$ also diverge. In addition the crossover on the metallic side is again characterized by two scales. The marginal Fermi liquid regime of Fig. \ref{mtttrsn} is replaced by a non-fermi liquid metal which evolves into the Landau Fermi liquid at the lowest energies.

We also note an earlier study\cite{Hermele} of the Mott transition in a half-filled Hubbard model on the two dimensional honeycomb lattice to a spin liquid phase. As is well known the conducting phase on this lattice consists of Dirac points and not a full Fermi surface. The spin liquid phase considered in that work also inherits this Dirac structure. This is a big difference from the Fermi surface case studied in this paper, and leads to very different results. Another recent study\cite{qiss} discussed the evolution of gapped spin liquid insulators into the usual Fermi liquid on a triangular lattice and suggested that this proceeds through various exotic lattice symmetry broken states. In this paper we instead focus on the transition to the gapless spin liquid state from the metal.

The rest of the paper is organized as follows. We begin in Section \ref{model} by defining the model and the slave rotor field theory that we employ to study the Mott transition. We then analyse the field theory at the $T = 0$ Mott transition and show that it could stay second order beyond mean field. We argue that the gauge field does not affect the critical properties of the bosons, and determine the structure of the strongly coupled spinon-gauge sector system. Next in Section \ref{fls} we study the approach to the Mott transition at $T = 0$ from the Fermi liquid side. Critical singularities of several properties are determined.
Next in Section \ref{thermo} we briefly study the singularities in thermodynamic quantities at the critical point itself. We then turn to transport properties in Section \ref{trsprt} demonstrating in particular the universal resistivity jump discussed above. In Section \ref{cfs} we show that the ground state at the critical point is characterized by a sharp Fermi surface but no Landau quasiparticle. Next in Section \ref{motts} we briefly discuss the approach to the Mott transition at $T = 0$ from the insulating side. The structure of the full crossover from the critical point to either phase is considered next in Section \ref{mflsection}. Here we show the existence of two energy scales characterizing the crossover, and demonstrate the emergence of marginal fermi liquids. The filling controlled transition is briefly studied in Section \ref{fcmott}. Possible implications for experiments particularly in $\kappa-(ET)_2Cu_2(CN)_3$ are discussed in Section \ref{expts}. We conclude with a brief discussion of some theoretical issues. Two appendices provide some details of the calculation of the spinon self energy and the electron vertex.

\section{Model and field theory}
\label{model}
We consider a one band Hubbard model at half-filling on a non-bipartite lattice such as the triangular lattice:
\begin{equation}
H = -t\sum_{<rr'>} \left(c^\dagger_{r}c_{r'} + h.c \right) + U\sum_r \left(n_r - 1\right)^2
\end{equation}
where $c_r$ destroys a spinful electron at site $r$ of a triangular lattice. $n_r = c^\dagger_rc_r$ is the electron number at site $r$. $U > 0$ is an on-cite repulsion. For large $g = t/U$ the ground state is a stable Fermi liquid metal. For small $t/U$ a Mott insulator results. Clearly there needs to be a Mott metal insulator transition at some critical value of $g_c = \left(t/U\right)_c$.

In application to electronic materials like
$\kappa-(ET)_2Cu_2(CN)_3$, it is important to include the long range part of the Coulomb interaction.
The Hubbard model can also possibly be realized in cold fermionic atoms in a periodic optical lattice in which case there is no long range Coulomb interaction. We begin by first studying the Mott transition in the absence of the Coulomb interaction.  At the end of the paper we discuss the necessary modifications if such Coulomb interaction is present.

The Mott transition and the spin liquid phase are conveniently discussed using the slave rotor representation of Ref. \onlinecite{florens}. We write
\begin{equation}
c_{r\alpha} = e^{i\phi_r} f_{r\alpha}
\end{equation}
with $e^{i\phi_r} \equiv b_r$ a spin-$0$ charge-$e$ boson, and $f_{r\alpha}$ a spin-$1/2$ charge-$0$ fermionic spinon. We start with a mean field description in which the spinons are non-interacting and form a Fermi surface. If the boson $b_r$ is condensed ($ \langle b_r \rangle \neq 0$) the result is the Fermi liquid phase of the electrons. If the boson is gapped (and hence uncondensed) a spin liquid Mott insulator with a spinon Fermi surface results. The phase transition at $g_c$ between the two phases is driven by the condensation of the boson $b_r$. A low energy effective theory\cite{leesq} for the transition is given by the action
\begin{eqnarray}
\label{mttS1}
S & = & S_b + S_f + S_a + S_{bf}\\
S_b & = & \int d^2x d\tau |\left(\partial_\mu - i a_\mu \right) b|^2 + V\left(|b|^2\right) \\
S_f & = & \int_{\vec x, \tau} \bar{f} \left(\partial_{\tau} - \mu_f + ia_0 \right)f + \int_{\vec k, \omega}\bar{f}_{\vec k, \omega}\epsilon^f_{\vec k +  \vec a}f_{\vec k, \omega} \\
S_a & = & \int_{\vec x, \tau}\frac{1}{e_0^2}\left(\epsilon_{\mu\nu\lambda} \partial_\nu a_\lambda \right)^2
\end{eqnarray}
The $a_{\mu}$ is a $U(1)$ gauge field that appears due to the redundancy introduced by the slave rotor representation of the electron operators. The potential $V\left(|b|^2\right)$ may simply be taken to be of the usual form $r|b|^2 + u|b|^4$. $\epsilon^f$ is the mean field spinon dispersion. The last term $S_{bf}$ represents residual short range interactions between the bosons and fermions. Potentially the most important of these is a coupling between $|b|^2$ and  long wavelength deformations of the spinon Fermi surface\cite{morin}.

Before proceeding we note that in the microscopic derivation of this action\cite{leesq} the $U(1)$ gauge field $a_{\mu}$ should be taken to be compact. However as in the theory of gapless spin liquids with a large number of gapless matter fields\cite{stableu1} the spinon Fermi surface is expected to supress space-time monopole configurations of the gauge field at low energies. Therefore we will henceforth take the gauge field $a_{\mu}$ to be non-compact.

In the `mean field' approximation we ignore the gauge fields but not other interactions. The boson condensation transition of action $S_b$ is then in the $3D$ XY universality class. Let us now consider the effect of the interaction terms $S_{bf}$. The operator $O = |b|^2$ can couple to the particle-hole continuum of the spinon Fermi surface near $q = 0$. The arguments of Ref. \onlinecite{morin} now show that this is an irrelevant perturbation at the $3D$ XY fixed point. Integrating out the spinon Fermi surface leads to a perturbation of the form\cite{morin}
\begin{equation}
\label{bbff}
v\int_{\omega,\vec q} \frac{|\omega|}{q} |O(\vec q, \omega)|^2
\end{equation}
The operator $O$ has scaling dimension $3 - \frac{1}{\nu}$ at the $3D$ XY fixed point where $\nu$ is the correlation length exponent. Thus $v$ can be seen to irrelevant so long as $\nu > \frac{2}{3}$. This inequality is satisfied for the $3D$ XY model.

Thus the bosons and spinons are decoupled in the absence of gauge interactions.  The electron Green's function in the Fermi liquid side is then simply given by
\begin{equation}
G^{(mf)}_c(\vec K, \omega) \approx |\langle b \rangle|^2 G^{(mf)}(\vec K, \omega)
\end{equation}
where $G^{(mf)}_{c,f}$ are the electron and spinon Greens functions within this mean field approximation.
This leads to a quasiparticle  residue
$Z \sim |\langle b \rangle|^2$. Clearly $Z$ vanishes on approaching the Mott transition due to the vanishing condensate fraction
as $Z \sim |g - g_c|^{2\beta}$ where $\beta$ is the order parameter exponent for the $3D$ XY model.
In this approximation the quasiparticle effective mass does not diverge and stays finite as the transition is approached.  In Ref. \onlinecite{critfs} we calculated the electron spectral function at the critical point  and found the scaling form
\begin{equation}
A_c^{(mf)}(\vec K, \omega) \sim |\omega|^{\eta} F^{(mf)}\left(\frac{c_0 \omega}{k_\parallel}\right)
\end{equation}
Thus in the simple mean field calculation there is  a sharp critical Fermi surface at which the electron spectral function has singularities even though there is no Landau quasiparticle.

Now we consider how these results are modified upon including gauge fluctuations. We begin by treating the gauge field within the standard Random Phase Approximation (RPA). As in other problems with a fermi surface coupled to the gauge field this is actually expected to capture the exact low energy form of the gauge propagator even beyond the RPA. We thus expect that the critical properties calculated below are a correct description of the low energy physics.
We work in the Coulomb gauge so that the $a_0$ and the transverse spatial component of the gauge field $\vec a$ are decoupled.
The $a_0$ component is screened out by the finite density of spinons, and can be integrated out.    the effective action for the transverse gauge field found by integrating out all matter fields and truncating to quadratic order is
\begin{equation}
S_{eff}[a] = \int_{\vec q,\omega} \left(\frac{k_0 |\omega|}{ q} + \chi_d q^2 + \sigma_0 \sqrt{\omega^2 + c^2 q^2}\right)|a(\vec q, \omega)|^2
\end{equation}
The first two terms come from the spinons and the third term from the bosons. The fermion contribution is well known and involves the important Landau damping term $\frac{k_0 |\omega|}{ q}$ with
$k_0$ of order the typical spinon Fermi momentum. The $\chi_d$ is the diamagnetic susceptibility of the spinons. The
the boson contribution is what obtains right at the critical point, and reflects the presence of a non-zero universal boson conductivity $\sigma_0$ at the mean field boson fixed point. $c$ is the space-time rescaling factor in the boson sector. Away from the critical point this boson contribution will be modified (see Section \ref{fls}).

In the presence of the $\frac{|\omega|}{|q|}$ term we expect that the important gauge fluctuations will involve frequencies $|\omega| \ll q$. Then in the low $|\omega|,q$ limit the gauge action may be approximated to
\begin{equation}
\label{Sa}
S_{eff}[a] = \int_{\vec q,\omega} \left(\frac{k_0 |\omega|}{v_{F0} q} +  \sigma_0 cq  \right)|a(\vec q, \omega)|^2
\end{equation}
This gives overdamped dynamics for the gauge field. The same structure of gauge field dynamics is also encountered in the theory of the half-filled Landau level\cite{hlr} and we may directly take over many results. First the spinon self energy at zero temperature due to interaction with this gauge field
is given by
\begin{equation}
\label{fself}
\Sigma_f(\vec K, i\omega) = v_{F0}^2 \int_{\vec q, \Omega} \left(\frac{1}{\frac{k_0 |\Omega|}{v_{F0}q}+\sigma_0 c |q|}\right)\frac{1}{\left(i\Omega - \epsilon^f_{\vec K - \vec q}\right)}
\end{equation}
Here $\epsilon^f$ is the mean field spinon dispersion. We are interested in this integral for $\vec K$ close to the spinon Fermi surface and small $\omega$. From Ref. \onlinecite{hlr} we have (see also Appendix \ref{selfapp})
\begin{equation}
\label{mslself}
\Sigma_f (\vec K, i\omega) \sim ia\omega ln\left(\frac{1}{|\omega|}\right)
\end{equation}
with $a$ a  non-universal constant.

The boson self energy due to the interaction with the gauge field is also readily evaluated.  To leading order the $(\vec k, \omega)$ dependence of the self energy is given by
\begin{equation}
\Sigma_b(\vec k, i\omega) - \Sigma_b(0,0)  =  \int_{\vec q, \Omega} \frac{\left(\vec k \times \hat{q} \right)^2}{\frac{k_0 |\Omega|}{v_{F0}q}+\sigma_0 c |q|}{\cal G}^0_b(\vec k - \vec q, \omega - \Omega)
\end{equation}
where ${\cal G}^0_b$ is the boson propagator at the mean field critical point:
\begin{equation}
{\cal G}^0_b(\vec q, \Omega) = \frac{1}{\left(\Omega^2 + c^2q^2\right)^{\frac{2-\eta}{2}}}
\end{equation}
Here $\eta$ is the anamolous dimension of the boson field at  the $3D$ XY fixed point. ($\Sigma_b(\vec k = 0, i\omega = 0)$ leads to a trivial shift of the location of the critical point and hence has been subtracted out).
Evaluating the integral we find
 \begin{equation}
\Sigma_b(\vec k, i\omega) - \Sigma_b(0,0) \sim k^2
\end{equation}
in the limit $k \rightarrow 0, |\omega| = 0$ limit, and is zero if $\vec k = 0, \omega \rightarrow 0$.  We see that the gauge interaction only leads to an analytic correction to the inverse boson propagator and hence does not alter the critical singularities coming from the boson self-interaction.

In fact the gauge coupling does not at all alter the critical properties of the bosons. At the mean field critical fixed point the bosons have dynamical critical exponent $1$ so that $\omega$ and $\vec q$ should scale identically. With that scaling the $\frac{\omega}{q}$ term in the gauge field action scales in the same way as a Higgs mass term. This term quenches the effects of the gauge fluctuations in the boson sector. Thus as far as the boson sector is concerned the critical properties are the same as the $3D$ XY fixed point.

In contrast in thinking about the effects of the gauge fluctuations on the fermions the scaling is different, as has been emphasized in various prior treatments of the problem of a Fermi surface coupled to a gauge field\cite{leenag,hlr,polch,aliomil,chetan}. Consider any definite point on the Fermi surface with, say normal along the $\hat{x}$ direction and let the momentum deviation from this point be $\vec q = (q_x, q_y)$. Then $q_x$ scales like the frequency $\omega$ but the tangential momentum scales as $q_y^2 \sim \omega$. The quadratic scaling of frequency with the tangential momentum means that gauge fluctuations (which from Eqn. \ref{Sa} also have such quadratic scaling)  can couple efficiently by transferring tangential momentum to the fermions.

Thus we have the interesting situation that the bosonic sector behaves as it would in the absence of any gauge field while the spinon sector is modified (albeit in well understood ways) by the gauge field.
This enables us to reliably analyse the critical point in great detail.

Note that the argument for the irrelevance of $S_{bf}$ remains unmodified even after taking into account the modification of the fermion sector by the gauge field. This is because even with the gauge field present the two particle Greens functions of the spinons at small $q$ retain their `Fermi liquid' form\cite{ybkim}, and hence the term in Eqn. \ref{bbff} is unmodified.

Before proceeding we note one other result we will need. The physical electron $c_{\alpha}$ is a product of the boson and fermion operators. At mean field level the electron Green's function is simply a convolution of the boson and fermion Green's functions. Beyond mean field, apart from the self-energies just discussed, we also need the correction to the $c_{\alpha} \rightarrow b + f_\alpha$ vertex from gauge fluctuations. In Appendix \ref{vertexapp} we show that (at $T = 0$) there is no singular enhancement of this vertex at low frequency/momenta near the Fermi surface. Therefore we can simply calculate the electron Greens function by convolving the boson and fermion Greens functions.

\section{Approach from the Fermi liquid}
\label{fls}
We first consider the approach to the Mott transition from the Fermi liquid side. In terms of the $b$ and $f_\alpha$ this phase is the boson condensate $\langle b \rangle \neq 0$. This has the immediate effect that the boson contribution to the gauge propagator is modified. At the longest wavelengths the gauge field acquires a `Higgs' mass proportional to the phase stiffness $\rho_s$ of the condensed boson. In general we write
\begin{equation}
\label{safl}
S_{eff}[a] = \int_{\vec q, \omega} \left(\frac{k_0 |\omega|}{ q} + \chi_d q^2 +  \Pi_b\left(\vec q, \omega, \rho_s\right)\right)|a(\vec q, \omega)|^2
\end{equation}
where the boson polarizability $\Pi_b$ satisfies the scaling form
\begin{equation}
\Pi_b(\vec q, \omega, \rho_s) = \sigma_0\sqrt{c^2q^2 + \omega^2}P\left(\frac{\sqrt{c^2q^2 + \omega^2}}{\rho_s}\right).
\end{equation}
The universal scaling function $P(x)$ behaves as
\begin{eqnarray}
P(x \rightarrow 0) & \sim & \frac{1}{x} \\
P(x \rightarrow \infty) & \sim & 1
\end{eqnarray}
The boson phase stiffness $\rho_s$ goes to zero at the transition as the inverse correlation length
\begin{eqnarray}
\rho_s & \sim & \frac{1}{\xi} \\
\xi & \sim & |g-g_c|^{-\nu}
\end{eqnarray}
where $\nu$ is the correlation length exponent of the $3D$ XY model.

Repeating the calculation of the fermion self energy in Eqn. \ref{fself} with this modified gauge propagator we find (see Appendix \ref{selfapp})
\begin{equation}
\label{sigmaffl}
\Sigma_f(\vec K, i\omega) =  2ia\omega ln\frac{1}{\rho_s} + o(\omega^2)
\end{equation}
Thus the finite gauge field mass $\rho_s$ cuts off the frequency dependent logarithm in the self energy
that obtains at the critical point.   The electron Greens function close to the Fermi surface is then simply determined to be
\begin{equation}
\label{gcfl}
{\cal G}_c(\vec K, \omega) = \frac{|\langle b \rangle|^2}{i\omega\left(1 + 2aln\frac{1}{\rho_s}\right) - v_{F0} k_{\|}}
\end{equation}
where $k_{\|}$ is the deviation from the Fermi momentum in the direction parallel to the normal to the Fermi surface. For small $\rho_s$ this gives
\begin{equation}
{\cal G}_c(\vec K, \omega) = \frac{Z}{i\omega - v_F k_{\|}}
\end{equation}
The quasiparticle residue $Z$ behaves as
\begin{eqnarray}
\label{Zs}
Z & \sim & \frac{|\langle b \rangle|^2}{ln\frac{1}{\rho_s}} \\
& \sim & \frac{|g-g_c|^{2\beta}}{ln\frac{1}{|g - g_c|}}
\end{eqnarray}
where $\beta$ is the order parameter exponent for the $3D$ XY model. Thus $Z$ vanishes as the Mott transition is approached. The power law dependence on $|g-g_c|$ is already obtained within mean field
theory; gauge fluctuations lead to the extra logarithm in the denominator. The renormalized Fermi velocity behaves as
\begin{equation}
\label{effms}
\frac{v_F}{v_{F0}} \sim \frac{1}{ln\frac{1}{|g-g_c|}}
\end{equation}
Thus the quasiparticle effective mass diverges logarithmically on approaching the Mott transition. This divergence is entirely a fluctuation effect and is absent in a mean field treatment.

In the Fermi liquid the specific heat $C_v \sim \gamma T$, and the divergence of the effective mass immediately implies that
\begin{equation}
\gamma \sim ln\frac{1}{|g-g_c|}
\end{equation}
close to the transition.

In the problem of a spinon Fermi surface coupled to a gauge field the spin susceptibility on the other hand is known not to be enhanced by the gauge interactions. Thus $\chi_0$ goes to a finite non-zero constant as the transition is approached. (Indeed it continues to a finite non-zero value in the insulating spin liquid as well). In Fermi liquid theory
\begin{equation}
\chi_0 \sim \frac{\rho_o}{1 + F^a_0}
\end{equation}
Here $\rho_0 \sim \int_{FS} \frac{1}{v_F}$ is the quasiparticle density of states at the Fermi surface and $F^a_0$ is a Landau parameter. As $\rho_0$ diverges while $\chi_0$ stays constant we infer that the Landau parameter $F^a_0$ diverges in exactly the same way as the effective mass:
\begin{equation}
F^a_0 \sim ln\frac{1}{|g-g_c|}
\end{equation}

Next we consider the compressibility $\kappa = \frac{dn}{d\mu}$ where $n$ is the electron density and $\mu$ is the electron chemical potential. This quantity receives contributions from both the bosons and the fermions. Indeed there is an Ioffe-Larkin composition rule\cite{ilcomp}:
\begin{equation}
\kappa^{-1} = \kappa_b^{-1} + \kappa_f^{-1}
\end{equation}
where $\kappa_{b,f}$ are the compressibilities of the boson and spinon subsystems. The boson compressibility vanishes as the Mott transition is approached in a well known way. We have
\begin{equation}
\kappa_b \sim \frac{1}{\xi}
\end{equation}
The spinon system on the other hand has a finite non-zero compressibility which (like the spin susceptibility) is not enhanced by gauge fluctuations. Clearly then the full compressibility is dominated by the bosons and vanishes at the transition. We have
\begin{eqnarray}
\kappa & \sim & \frac{1}{\xi} \\
& \sim & |g-g_c|^{\nu}
\end{eqnarray}

In Fermi liquid theory the compressibility may be written as
\begin{equation}
\kappa \sim \frac{\rho_o}{1 + F^s_0}
\end{equation}
where $F^s_0$ is a different Landau parameter. The combination of diverging effective mass and vanishing compressibility implies a strongly diverging Landau parameter:
\begin{equation}
F^s_0 \sim \xi ln\xi
\end{equation}
We note that the diverging $F^0_s$ implies that the zero sound velocity will diverge.

If we had kept the term of $o(\omega^2)$ in Eqn. \ref{sigmaffl} we would have found the usual quasiparticle decay rate $\gamma \propto \omega^2$. Including it we find that
\begin{eqnarray}
\gamma & \sim & \frac{\omega^2}{\rho_s^2} \\
& \sim & \xi^2 \omega^2 \\
& \sim & |g-g_c|^{-2\nu} \omega^2
\end{eqnarray}

To summarize the approach to the Mott transition from the Fermi liquid is characterized by a vanishing $Z$, diverging effective mass, and diverging $F^{s,a}_0$ (such that the spin susceptibility stays constant and the compressibility vanishes). We emphasize that even though $Z$ vanishes and the effective mass diverges, $Z$ is not inversely proportional to the effective mass (a result that is even more strikingly true at the mean field level where there is vanishing $Z$ but no mass divergence). This is in striking contrast to results from Dynamical Mean Field Theory for the Hubbard model in infinite dimensions\cite{dmft}.

\section{Thermodynamic singularities at the critical point}
\label{thermo}
Now we briefly consider singularities in thermodynamic properties right at the Mott critical point. The specific heat receives contributions from both the critical bosons and the coupled spinon-gauge system. The former contribution
is well known and goes like $T^2$ at low-$T$. The spinon-gauge contribution is readily calculated and behaves as
\begin{equation}
C_v \sim T ln \left(\frac{1}{T}\right)
\end{equation}
This is therefore the dominant contribution at the critical point.

The compressibility at a non-zero $T$ at the critical point is determined by the Ioffe-Larkin rules.
The boson compressibility $\kappa_b \sim \frac{1}{T}$ while $\kappa_f$ is a temperature independent constant. Therefore at low $T$ the boson contribution dominates and we find
\begin{equation}
\kappa \sim \frac{1}{T}
\end{equation}

The spin susceptibility is of course determined by the spinons and is a $T$-independent constant at low $T$.

\section{Transport}
\label{trsprt}
We now consider the transport properties. We will assume that there is some weak disorder which gives some non-zero residual resistivity in the metal. In principle if the associated elastic mean free path is $l$ the Landau damping term in the gauge action will be modified from $\frac{|\omega|}{q}$ to $|\omega|l$ for $q \ll \frac{1}{l}$. We will
assume that $l$ is large enough so that this modification can be ignored at all but the lowest temperatures. With this caveat we now consider d.c transport across the Mott transition.

The resistivity $\rho$ is again determined by the Ioffe-Larkin rule which states that
\begin{equation}
\rho = \rho_b + \rho_f
\end{equation}
Consider first the residual resistivity at $T = 0$ ({\em i.e} the resistivity extrapolated from  temperatures higher than the low temperature at which the modification of the gauge propagator due to disorder must be considered). In the Fermi liquid phase the bosons are condensed and so $\rho_b = 0$. We therefore get
$\rho = \rho_f$. In the Mott insulator on the other hand $\rho_b = \infty$ and hence $\rho = \infty$ at it must be. Thus the $T = 0$ conductivity jumps on crossing the Mott transition. A similar argument was first used by Coleman et. al.\cite{cms} to argue that the residual resistivity jumps at the Kondo breakdown transition of Ref. \onlinecite{svsrev}. In the present problem however there is a new interesting feature associated with the resistivity jump which is absent at the Kondo breakdown transition. Consider the resistivity right at the quantum critical point. Then from prior studies of the superconductor-insulator transition of bosons\cite{fgg} we know that $\rho_b = \frac{Rh}{e^2}$ is non-zero and {\em universal}. Thus the residual resistivity $\rho_{cr}$ at the critical point is
\begin{equation}
\rho_{cr} = \frac{Rh}{e^2} + \rho_f
\end{equation}
Thus the residual resistivity right at the critical point jumps from its value in both the metallic phase (where it is $\rho_f$) and the insulating phase (where it is $\infty$). Furthermore we see that on approaching from the metallic side the jump in the resistivity at the critical point is the universal number $\frac{Rh}{e^2}$. This remarkable {\em universal resistivity jump} can be tested in experiments.

It is instructive to consider the behavior of the conductivity $\sigma = \frac{1}{\rho}$ as we move across the transition. This is depicted in Fig. \ref{Rjmp}. The conductivity jumps twice and takes a value at the critical point intermediate between that of the metal and the insulator. However clearly neither conductivity jump is universal.

The universal jump in resistivity rather than the conductivity is closely tied to the Ioffe-Larkin rule. Thus observation of the universal resistivity jump can provide direct support for the Ioffe-Larkin rule, and hence the general correctness of the gauge theoretic description of the transition.

Now lets consider the $T$ dependence of the resistivity at the critical point. The spinon conductivity is determined by the scattering off the critical gauge fluctuations. The calculation of the transport lifetime of the spinons is standard and is described in Ref. \onlinecite{leenag} for the case where the $\omega$-independent term in the gauge propagator is $q^2$ (as opposed to the $|q|$ that arises in the present problem). The transport scattering rate at low temperature is thus given by
\begin{eqnarray}
\gamma_{tr} & \sim & \int_0^T d\omega \frac{T}{\omega}\int dq \frac{\omega q^3}{\omega^2 + \left(\frac{\sigma_0c q^2}{k_0}\right)^2} \\
& \sim & T^2 ln\left(\frac{1}{T}\right)
\end{eqnarray}
Thus the spinon resistivity will have a $T^2ln\left(\frac{1}{T}\right)$ dependence right at the quantum critical point. The boson resistivity is only expected to have weak temperature dependent corrections to its universal $T = 0$ value so that the leading temperature dependence of the total resistivity is $T^2$ with a log correction.

Finally let us consider the thermal conductivity. In the metallic phase the Wiedemann-Franz law is obeyed for the residual $T \rightarrow 0$ conductivity so that the thermal conductivity $K_m = L\sigma_m T$ where $L = \frac{\pi^2}{3}\left(\frac{k_B}{e}\right)^2$ is the usual Lorenz number. What happens right at the quantum critical point? In contrast to electrical transport the thermal conductivities of the boson and spinon-gauge systems add\cite{leenag}. The residual thermal conductivity of the spinon-gauge system is expected to approximately satisfy the Wiedemann-Franz law and be of order $K_m$. However in the scaling limit at the critical point the relativistic invariance of the boson system implies that the thermal conductivity is actually infinite\cite{rqcthrmcond}. The boson thermal conductivity will then be determined by perturbations that are formally irrelevant at the critical point. These will lead to a large  thermal conductivity at low temperature. Thus we expect that the boson contribution will dominate over the spinons, and that the Wiedemann-Franz law will be violated at the quantum critical point.
Low temperature thermal conductivity in the insulating phase has been discussed in Ref. \onlinecite{leesq}, and goes like $T^{2/3}$ coming entirely from the spinons.

\section{Critical Fermi surface}
\label{cfs}

In Ref. \onlinecite{critfs} we argued that at a second order Mott transition, the Fermi surface will continue to be sharply defined even though the Landau quasiparticle is absent. We dubbed this a `critical Fermi surface'. We also demonstrated the existence of such a critical fermi surface within the slave rotor mean field theory of the Mott critical point discussed in this paper. Here we show that the critical Fermi surface remains sharply defined even beyond the mean field approximation.

To leading order the $c_\alpha \rightarrow b+f_\alpha$ vertex is not singularly enhanced by gauge fluctuations. Therefore
the electron Greens function ${\cal G}_c(\vec x, \tau)$ is simply the product of the boson and spinon Greens functions:
\begin{equation}
{\cal G}_c(\vec x, \tau) = {\cal G}_b(\vec x, \tau){\cal G}_f(\vec x, \tau)
\end{equation}
The electron spectral function $A_c(\vec K, \omega)$ for real positive frequencies is then given by
\begin{equation}
\label{convolve}
A_c(\vec K, \omega) = \int_{\vec q}\int_0^\omega d\Omega A_b(\vec q, \Omega) A_f\left(\vec K - \vec q, \omega - \Omega\right)
\end{equation}
with $A_{b,f}$ the boson and spinon spectral functions respectively. At the critical point of interest these
take the form
\begin{eqnarray}
A_b(\vec q, \Omega) & = & {\cal A}  \frac{\theta\left(\Omega^2 - c^2q^2\right)}{\left(\Omega^2 - c^2 q^2\right)^{\frac{2-\eta}{2}}} \\
A_f(\vec q, \Omega) & = &  \frac{\frac{\pi \gamma \Omega}{2}}{\left(\frac{\pi \gamma \Omega}{2}\right)^2
+ \left(\Omega ln\frac{\Lambda}{\Omega} - \epsilon^f_q \right)^2}
\end{eqnarray}
The spinon Greens function includes the non-trivial self energy that arises from the gauge field interaction. ${\cal A}, \Lambda, \gamma$ are all non-universal constants. Let us consider a momentum $K = \left(K_F + k\right)\hat{x}$ with $k$ small. The spinon is at momentum
$\vec K - \vec q$. The spinon energy may be taken to be
\begin{equation}
\epsilon^f_{\vec K - \vec q} \approx v_{F0}(k - q_x) + Cq_y^2
\end{equation}
with $C$ related to the curvature of the spinon Fermi surface. Putting these into
Eqn. \ref{convolve},   for small $|k|, \omega$ the important region of integration involves $|q_x| \sim |q_y| \sim \Omega \sim \omega$. Thus we may drop the curvature term $Cq_y^2$ in the fermion dispersion above. Further
at the lowest frequencies $\Omega ln\frac{\Lambda}{\Omega} \gg \Omega$ so that we may drop the imaginary part of the spinon self energy compared to its real part. Note also that the momentum transfer $|q_x| \sim \omega \ll \omega ln\frac{1}{\omega}$ and hence to leading order we should drop that as well. The spinon spectral function can then be replaced with a delta function:
\begin{equation}
A_f(\Omega, \vec K - \vec q) \approx \delta\left(\Omega ln\frac{\Lambda}{\Omega} - v_{F0}k\right)
\end{equation}
The $q$ integrals in eqn. \ref{convolve} may now be readily performed and we find
\begin{equation}
A(\vec K, \omega) \sim \int_0^\omega d\Omega \left(\omega - \Omega\right)^\eta \delta\left(\Omega ln\frac{\Lambda}{\Omega} - v_{F0}k\right)
\end{equation}
A scaling limit can now be defined with scaling parameter $\frac{\omega ln\frac{\Lambda}{\omega}}{v_{F0}k}$. In the limit that $k, \omega$ go to zero while keeping this parameter constant, we find
\begin{equation}
A(\vec K, \omega) \sim \frac{\omega^\eta}{ln\frac{\Lambda}{\omega}}F\left(\frac{\omega ln\frac{\Lambda}{\omega}}{v_{F0}k}\right)
\end{equation}
with the scaling function
\begin{equation}
F(x) = \left(1 - \frac{1}{x}\right)^\eta \theta(x - 1)
\end{equation}

Thus we conclude that the electron spectral function has sharp singularities at the Fermi surface  right at the critical point even after including gauge fluctuations. However the Landau quasiparticle is absent.
Thus this provides an explicit example of a `critical Fermi surface' at the quantum critical point.

\section{Approach from the insulating side}
\label{motts}
Now we briefly consider approaching the Mott transition at $T = 0$ from the insulating side. First clearly there will be a charge gap determined by the boson gap $\Delta$ which vanishes as
\begin{equation}
\Delta \sim \frac{1}{\xi}
\end{equation}
We emphasize that the charge gap vanishes at the same point that the Fermi surface disappears on approaching from the metallic side, once again in contrast to results in infinite dimension\cite{dmft}. This was already pointed out at the mean field level\cite{florens}, and continues to be the case after including fluctuations.

The electron spectral function is of course also gapped in the Mott insulator. This gap derives from the boson gap $\Delta$ as the spinon is gapless even in the insulator. On approaching the transition this single particle gap vanishes at the momenta that correspond to the Fermi surface of the metallic phase and not at isolated momenta. This is exactly as expected for a continuous Mott transition\cite{critfs}.

The Mott insulating state is a gapless quantum spin liquid with a spinon Fermi surface. This is a non-fermi liquid state of the spinons with well understood properties\cite{leenag,lesik,leesq,aliomil}. For instance the low-$T$ specific heat goes like $T^{2/3}$. The proportionality coefficient will vanish as the critical point is approached in a manner that is readily determined.
In the insulator the gauge propagator will be given by a form analogous to Eqn. \ref{safl} except that the $\rho_s$ that appears in the scaling function $P$ is replaced by $\Delta$. At small momenta $q$ the boson polarizability must go as $\frac{q^2}{\Delta}$. This implies that the $T^{2/3}$ specific heat must have a coefficient $D$ that vanishes as
\begin{equation}
D \sim \Delta^{\frac{1}{3}}ln\frac{1}{\Delta}
\end{equation}

\section{Crossover out of criticality: Marginal Fermi Liquids}
\label{mflsection}
In this section we study in greater detail the crossover out of criticality.
Consider first the boson sector. On moving away from the critical point the crossover out of criticality is determined by a single energy scale that vanishes at the critical point. On the metallic side this scale may be taken to be the boson stiffness $\rho_s$, and on the insulating side the boson gap $\Delta$.
Let us focus on the metal. The structure of the gauge propagator Eqn. \ref{safl} implies that it is modified from the critical point at momenta $q \sim \frac{\rho_s}{c}$ or a frequency $\omega \sim \frac{\rho_s^2}{ck_0}$. Thus the frequency scale at which the gauge fluctuations notice the Bose condensation is much smaller than the scale $\rho_s$. This can also be seen in the spinon self energy.  In Appendix \ref{selfapp} we show that this can be written as
\begin{equation}
\label{sigmadiff}
\Sigma_f(\omega;\rho_s) - \Sigma_f(\omega; \rho_s = 0) \sim ia\omega g\left(\frac{ck_0|\omega|}{\rho_s^2}\right)
\end{equation}
where the function $g(x) \sim ln x$ for small $x$ and goes to zero for large $x$. Thus we see that the spinon self energy is modified from its critical form at a scale $\sim \rho_s^2$ much smaller than the Bose condensation scale $\rho_s$.

What are the properties of the system in the intermediate energy range between $\rho_s$ and $\frac{\rho_s^2}{ck_0}$?
In that regime we can treat the bosons as having already condensed. However this condensation is not yet felt as a `Higgs' effect by the gauge fields, and hence by the spinons. Thus the coupled spinon-gauge system continues to behave as it would right at the quantum critical point. The electron Greens function in this regime will therefore have the same form as Eqn. \ref{gcfl} but with the $\rho_s$ in the spinon self energy replaced by $\sqrt{\omega}$. Thus we have
\begin{equation}
\label{gcmfl}
{\cal G}_c(\vec K, \omega) = \frac{Z_{MFL}}{i\omega\left(1 + aln\frac{1}{|\omega|}\right) - v_{F0} k_{\|}}
\end{equation}
with $Z_{MFL} = |\langle b \rangle |^2 \sim |g - g_c|^{2\beta}$. Remarkably this is exactly the same form of the electron Greens function postulated in the `Marginal Fermi Liquid' (MFL) state introduced by Varma et al\cite{varma} to describe the optimally doped cuprates.

At a non-zero temperature $T$ the two vanishing energy scales should manifest themselves as two distinct temperature scales $T^* \sim \rho_s$ and $T^{**} \sim \frac{\rho_s^2}{ck_0}$ that both vanish on approaching the transition. The properties are that of the quantum critical metal for $T \gg T^*$, that of the Marginal
Fermi Liquid for $ T^* \gg T \gg T^{**}$, and that of a Landau Fermi liquid for $ T \ll T^{**}$ (see Fig. \ref{mtttrsn}).

What are the properties of the marginal Fermi liquid state? Apart from the electron spectral function described above, it will have a specific heat $C(T) \sim T ln\frac{1}{T}$ unchanged from the quantum critical regime. This is because the specific heat is controlled by the spinon-gauge system which is still critical in the MFL. However the compressibility (which is dominated by the bosons) will be a temperature independent constant $\sim \frac{1}{\xi}$. The spin susceptibility will also be constant independent of temperature. Finally the resistivity will now be dominated by the fermions
and will extrapolate to a $T = 0$ value equal to just the spinon resistivity. This is again due to the Ioffe-Larkin rule. Below the Bose condensation scale $T^*$ the boson resistivity becomes very small so that the spinons give the dominant contribution.

Thus we see that the temperature scale at which the resistivity jump becomes evident is $T^*$. On the other hand the scale at which the crossover in the specific heat happens is $T^{**}$. The presence of the two energy scales thus leads to rich crossover phenomena near this transition.

Exactly the same considerations describe the insulating side as well. The boson gap $\Delta$ is not felt by the spinon-gauge system till a much lower energy scale $\frac{\Delta^2}{ck_0}$. In the intermediate temperature range the behavior is that of a spin liquid insulator with a marginal fermi liquid of spinons. We will call this a marginal spinon liquid to distinguish it from the non-fermi spinon liquid that obtains at the lowest temperatures. The marginal spinon liquid state is an incompressible Mott insulator with gapless spinon excitations at a Fermi surface (just like the low temperature spinon liquid of Refs. \onlinecite{lesik,leesq}). However it has a $T\ln\frac{1}{T}$ specific heat. Further the spinon scattering rate (given by the imaginary part of the spinon self energy) will be proportional to $T$; thus the thermal conductivity should behave as $\frac{K}{T} \sim \frac{1}{T}$. In contrast in the spinon non-fermi liquid state that obtains at low $T$ $\frac{K}{T} \sim \frac{1}{T^{2/3}}$.

Thus the primary transition that occurs is between a marginal Fermi liquid metal and a marginal spinon liquid Mott insulator. Both these states are eventually unstable at the lowest energies to the Landau Fermi liquid and the non-fermi spinon liquid state respectively. In renormalization group language we may say that the relevant flow away from the critical fixed point (say on the metallic side) leads to a marginal Fermi liquid fixed point. A different operator which is irrelevant at the critical fixed point is however relevant at the MFL fixed point, and that eventually leads to the Landau Fermi liquid fixed point. This is the classic instance of a `dangerously irrelevant' perturbation. Here it corresponds to the coupling between the bosons and the spinon-gauge system.

We note that the scaling hypotheses for the crossover out of criticality described in Ref. \onlinecite{critfs} presumed that the relevant flow out of the critical fixed point leads directly to the Landau Fermi liquid. As this does not happen in the present problem the approach to criticality at $T = 0$  should not be described by the scaling ansatz of Ref. \onlinecite{critfs}. This can be explicitly seen by considering for instance the scaling of the specific heat - the crossover scale is $T^{**}$ rather than the $T^*$ that would obtain if there was simple scaling. This feature may be a general limitation of gauge theoretic approaches to quantum phase transitions with a disappearing Fermi surface.

\section{Chemical potential tuned Mott transition}
\label{fcmott}
In this Section we very briefly consider the Mott transition that results when the spinon Fermi surface insulator is turned into a metal by the process of doping, {\em i,e} by tuning a chemical potential rather than by pressure at fixed filling. The results also describe the asymptotic critical behavior in the Kondo breakdown transition of a Kondo lattice studied in Ref. \onlinecite{svsrev}. The Kondo breakdown model has in addition to the sheet of the Fermi surface undergoing the Mott transition a separate Fermi surface sheet that is non-singular through the transition.  However Paul et al\cite{paul} have suggested that there may be a small but non-zero energy scale $E^*$ at the transition such that at temperatures above $E^*$ these additional sheets of the Fermi surface become important. In application to the Kondo breakdown model the results of this section only pertain to the asymptotic low energy regime $T \ll E^*$.

\begin{figure}
\includegraphics[width=8cm]{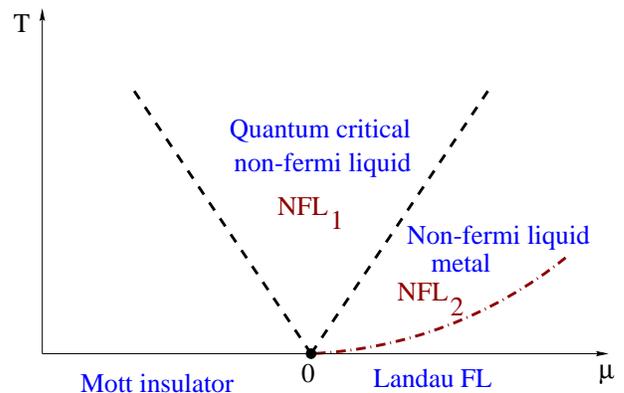}
\caption{Schematic phase diagram showing finite temperature crossovers near the filling controlled Mott transition in two dimensions. $\mu$ is the chemical potential. The Mott insulating ground state is a spin liquid with a spinon Fermi surface. The black dashed line is the crossover scale $T^*$ and the red dashed line is the crossover scale $T^{**}$. There are two non-fermi liquid regimes near the critical point denoted NFL$_1$ and NFL$_2$. The former appears right above the quantum critical point, and has weak singularities on a critical Fermi surface. $NFL_2$ appears between $T^*$ and $T^{**}$ on the metallic side and has strong singularities at the critical Fermi surface. The Landau Fermi liquid obtains only below $T^{**}$.} \label{fcmttxovr}
\end{figure}

A field theory appropriate for these transitions takes the same form as Eqn. \ref{mttS1} except that the boson action becomes non-relativistic.
\begin{equation}
S_b  =  \int d^2x d\tau \bar{b}\left(\partial_{\tau} - ia_0 - \frac{(\vec \nabla - i \vec a)^2}{2m_b} \right) b
\end{equation}
The physical electron operator is given by $c_\alpha = b f_\alpha$. Many properties of this transition were studied in Ref. \onlinecite{svsrev}. In particular it was found that the bosons decouple from the spinon-gauge system just like at the transition at fixed density studied in earlier sections in this paper. Furthermore the universal conductivity of the bosons at the chemical potential tuned transition is zero. Thus the gauge propagator retains the same structure that it has in the spin liquid phase itself. The physical properties in the quantum critical region are strikingly non-fermi liquid like. We will denote this non-fermi liquid NFL$_1$ to distinguish it from other non-fermi liquid states that appear in other regimes (see Fig. \ref{fcmttxovr}).
The electron Greens function at the critical point was not calculated in Ref. \onlinecite{svsrev}. This is easy to do as once again the $c_\alpha \rightarrow b + f_\alpha$ vertex is not singular at low energies so that the elecron Greens function may be obtained by convolution. The boson Greens function is
\begin{equation}
{\cal G}_b(\vec k, \omega) = \frac{1}{i\omega - \frac{k^2}{2m_b}}
\end{equation}
while the spinon Greens function is
\begin{equation}
{\cal G}_f(\vec k, \omega) = \frac{1}{i\lambda sgn(\omega) |\omega|^{\frac{2}{3}} - \epsilon^f_k}
\end{equation}
with $\lambda$ a non-universal constant. We find that there is a critical Fermi surface with the exponents $\alpha = -1, z = 2$. The large negative value of $\alpha$ means that there are weak singularities at this critical fermi surface.

Turning to the approach to the Mott transition from the Fermi liquid the gauge action now gets modified\cite{svsrev} to
\begin{equation}
\label{saflx}
S_{eff}[a] = \int_{\vec q, \omega} \left(\frac{k_0 |\omega|}{ q} + \chi_d q^2 +  \rho_s \right)|a(\vec q, \omega)|^2
\end{equation}
The divergence of the specific heat coefficient was calculated in Ref. \onlinecite{svsrev}. In two dimensions
\begin{equation}
\gamma \sim \frac{1}{\sqrt{\rho_s}} \sim \frac{1}{\sqrt{x}}
\end{equation}
where $x$ is the number density of doped electrons.  Calculating the spinon self energy due to interaction with gauge field we find
a spinon self energy $\Sigma_f \sim i\frac{\omega}{\sqrt{\rho_s}}$. This then gives an electron effective mass $\sim \frac{1}{\sqrt{\rho_s}}$ (in agreement with the $\gamma$ divergence) and an electron quasiparticle residue
\begin{eqnarray}
Z & \sim & |\langle b \rangle |^2 \rho_s \\
& \sim & x^{\frac{3}{2}}
\end{eqnarray}
Once again $Z$ is not inversely proportional to the effective mass. The spin susceptibility is again constant through the transition so that $F^0_a$ diverges exactly as the effective mass. The boson compressibility diverges logarithmically with $x$ and that determines the divergence of $F^0_s$.

Perhaps most interestingly the crossover out of criticality is again in two stages as with the bandwidth tuned transition (see Fig. \ref{fcmttxovr}). At a finite density $x$ the bosons condense below an energy scale $T^* \sim \rho_s$ but this is not felt by the spinon-gauge system till a lower energy scale $T^{**} \sim \rho_s^{\frac{3}{2}}$. In the intermediate energy regime the bosons may be treated as condensed but the spinon-gauge system has not yet emerged out of criticality. This leads to a genuine non-fermi liquid metal (NFL$_2$) with (at $T = 0$) an electron spectral function determined by the spinon Greens function (and an overall prefactor $\propto |\langle b \rangle |^2$). This state therefore has strong singularities at the critical Fermi surface. This non-fermi liquid metal will have a $T^{2/3}$ specific heat, a constant susceptibility, and a constant compressibility. It crosses over to the Landau Fermi liquid at low temperature. The arguments of Ref. \onlinecite{cms} show that the residual resistivity will jump to zero at $T = 0$ across the metal-insulator transition. At non-zero temperatures it is clear from our discussion that this jump will become evident across the $T^*$ line in the metallic phase.

\section{Experiments}
\label{expts}
Apart from its intrinsic theoretical interest the main motivation for this paper is to develop an understanding of the pressure tuned Mott transition in $\kappa-(ET)_2Cu_2(CN)_3$. In thinking about the experiments we first need to
dispose of some preliminaries. The theory developed in this paper was for a transition from a Fermi liquid metal to a Mott insulator with a spinon Fermi surface in two dimensions. In experiments on the metallic side superconductivity appears at low $T \sim 5 K$ near the Mott transition\cite{kanoda1}. As discussed in the Introduction there is evidence that a spin liquid state is indeed realized in the Mott insulator. This insulator also apparently has gapless spin excitations, thus supporting the proposal of Ref. \onlinecite{lesik,leesq} that it has a spinon Fermi surface.
However recent measurements\cite{etspht} of the low temperature specific heat have shown it to be linear in $T$ as opposed to the $T^{2/3}$ power law predicted for the spinon non-fermi liquid insulator\cite{lesik,leesq}. This could be due to a low temperature instability of the spinon fermi surface state\cite{amperean,kohn} that occurs at $T \sim 5 K$ and perhaps has the same origin as the superconductivity on the metallic side. The discussion in this paper applies to the physics above this low $T$ instability related to pairing.

We have thus far ignored the long range part of the Coulomb interaction. We can consider its effects by perturbing about the critical theory we have described in previous sections.  For the boson theory which describes the charge sector the long range Coulomb interaction is marginal by power counting\cite{fg}. An approximate renormalization group analysis was used to suggest that the transition could either be second order or driven first order by fluctuations
depending on parameters\cite{fg}. In the former case the Coulomb interaction is marginally irrelevant and will lead to log corrections to the properties we have calculated. At any rate we see that a second order Mott transition is allowed even after including the Coulomb interaction. Further except perhaps at the lowest temperatures, the criticality is unchanged from the short ranged case discussed in the bulk of the paper. Note that in the metallic phase the Coulomb interaction will get screened. The screening length will diverge on approaching the Mott transition in the same way as $\frac{1}{\sqrt{\kappa}}$ where $\kappa$ is the compressibility.

Now let us highlight some features of our results that directly bear on current or possible future experiments. First we note that even if the transition is second order the resistivity will jump. Thus observation of a resistivity jump is by itself not evidence of a first order transition. If the transition is second order our prediction of a universal resistivity jump of order $\frac{h}{e^2}$(see Fig. \ref{Rjmp}) should be observable. Note that this universal jump is in the in-plane sheet resistance of each layer. At non-zero $T$ the jump in resistivity will manifest itself as a strong crossover at the temperature scale $T^*$: on getting closer to the transition the evolution of the resistivity on crossing the $T^*$ line by varying pressure will get sharper. Other aspects of our results at the critical point (such as $Tln\frac{1}{T}$ specific heat and the Wiedemann-Franz violation) should also be observable.

It should also be interesting to map out the finite $T$ crossovers depicted in Fig. \ref{mtttrsn}. On the insulating side the crossover scale $T^* \sim \Delta$ the charge gap. The lower crossover temperature $T^{**}$ will then be of order $\frac{\Delta^2}{ck_0}$. The boson velocity $c$ can be roughly estimated as $\frac{\sqrt{Ut}a}{\hbar}$ in a Hubbard model description with $a$ the lattice parameter. As $k_0 \sim \frac{\hbar}{a}$ we estimate
\begin{equation}
T^{**} \approx \frac{\Delta^2}{\sqrt{Ut}}
\end{equation}
It is at present not clear from experiments precisely how big the charge gap is even at ambient pressure but it seems likely\cite{kanoda2} that it is much smaller than $t$ and $U$. Thus even for the ambient pressure material $T^{**}$ may be much smaller than $T^*$. It is interesting to consider the possibility (which will certainly be realized close to the transition) that the low $T$ instability discussed above occurs at a scale comparable to $T^{**}$. Then the crossover to the spinon non-fermi liquid insulator will not be observed, and much of the finite-$T$ physics will be that of the marginal spinon liquid state with properties as discussed in Section \ref{mflsection}.

The crossovers in the metallic side are perhaps of even greater interest. The presence  of an intermediate temperature state close to the Mott boundary raises the possibility that the metallic state above the superconducting transition is actually a marginal Fermi liquid and not a true Landau fermi liquid. Direct measurements of the single particle spectrum (perhaps by tunneling) will be thus be very useful.

\section{Discussion}
\label{disc}

In this paper we have developed a theory of a continuous Mott transition between a Fermi liquid and a paramagnetic Mott insulator with a spinon Fermi surface. As such this provides a valuable example of a quantum critical point where a Fermi surface disappears. Before concluding it is instructive to place our results in a more general context of
phase transitions involving the disappearance of a fermi surface. Such transitions were discussed from a general scaling point of view recently in Ref. \onlinecite{critfs}. First we see that this Mott critical point provides a concrete example of a sharp critical Fermi surface with no Landau quasiparticle. The scaling exponents at this surface are independent of position on the Fermi surface.  Even more importantly we found that the primary transition was not actually between the Fermi liquid state and the spinon Fermi surface ground state but rather between a marginal Fermi liquid state and its spinon counterpart. Thus universal scaling functions for the initial crossover out of criticality will only capture these marginal liquid phases. The physics of the Landau Fermi liquid metal is not obtained without accounting for the second crossover at a lower energy scale. Perhaps this is a feature of all slave particle approaches to such problems.

The slave particle approach is of course well suited to the particular transition studied in this paper (where the insulating state was a deconfined spin liquid). But there are other continuous phase transitions associated with the disappearance of a fermi surface where both phases are conventional. It is hardly clear that slave particle approaches are the way forward for such transitions. The scaling approach we developed in Ref. \onlinecite{critfs}
might provide guidance in searching for suitable alternate theoretical approaches.

\section*{Acknowledgments}
I thank  P.A. Lee, S. Sachdev, and R. Shankar  for useful discussions. This work was supported by NSF Grant DMR-0705255.
While this paper was being written I learnt that a field theory similar to the one in Section \ref{model} but with $2$ boson species is being studied by R. Kaul, S. Sachdev, and C. Xu\cite{ksx} in a different context. They conclude, as I do, that the criticality of the boson sector is unaffected by the gauge fluctuations.

\begin{appendix}
\section{Calculation of spinon self energy}
\label{selfapp}
First consider Eqn. \ref{fself} for the spinon self energy at the critical point. In evaluating that
integral we observe that due to the structure of the spinon propagator the important region of $q_x, q_y$ has
$q_x \sim q_y^2$. This implies that $|q_x| \ll |q_y$ so that the $q_x$ can be ignored compared to $q_y$ in the gauge propagator. The $q_x$ integral can now be done, and we get
\begin{equation}
\label{fslfcalc1}
\Sigma_f(\omega) = \frac{iv_{F0}}{2\pi} \int_{\Omega, q_y} \frac{sgn(\omega - \Omega)}{\frac
{k_0 |\Omega|}{|q_y|}+\sigma_0 c |q_y|}
\end{equation}
The $q_y$ integral is divergent in the ultraviolet. Imposing a cut-off $\Lambda$ we find (for small $\Omega$)
\begin{equation}
\Sigma_f = \frac{iv_{F0}}{4\pi^2 \sigma_0 c}\int_{\Omega} sgn(\omega - \Omega)ln \frac{\Lambda}{\Omega}
\end{equation}
The $\Omega$ integral can now be done and yields the result quoted in Eqn. \ref{mslself}.

Now consider the spinon self energy at low frequencies in the metallic phase. The necessary modification to the gauge propagator was
discussed in Section \ref{fls}. The self energy is then given by Eqn. \ref{fslfcalc1} but with the modified gauge propagator. The infrared cutoff for the $q_y$ integral is now set by $ \sim \rho_s^2$ rather than $\Omega$. Thus
we get Eqn. \ref{mslself} but with $\rho_s^2$ replacing $\omega$ for the argument of the logarithm.

Finally it is convenient to calculate the difference
$\Sigma_f(\omega;\rho_s) - \Sigma_f(\omega; \rho_s = 0)$. The advantage is that the $q_y$ integral that enters this
quantity is both ultraviolet and infrared convergent. So we can scale $q_y = y\rho_s$ and write the answer in the form \begin{eqnarray*}
\Sigma_f(\omega;\rho_s) - \Sigma_f(\omega; \rho_s = 0) & = & i\int_{\Omega} sgn(\omega - \Omega) \tilde{g}\left(\frac{ck_0|\omega|}{\rho_s^2}\right) \\
& = & ia\omega g\left(\frac{ck_0|\omega|}{\rho_s^2}\right)
\end{eqnarray*}
Thus the crossover from the critical spinon self energy to one characteristic of the Fermi liquid phase happens at an energy scale $\sim \rho_s^2$.

\section{Calculation of electron vertex}
\label{vertexapp}
Here we consider the correction to the $c_\alpha \rightarrow b+ f_\alpha$ vertex at $T = 0$ coming from the exchange of one gauge boson. Let $(\vec K, \vec K_f, \vec k_b)$ be the momenta of the electron, the spinon and the boson respectively and $(\omega, \omega_f, \omega_b)$ the corresponding frequencies. The vertex may be written
\begin{equation}
V\left(\vec K, \vec K_f, \vec k_b; \omega, \omega_f, \omega_b \right)\delta(\vec K - \vec K_f - \vec k_b) \delta(\omega - \omega_f -\omega_b)
\end{equation}
In the absence of the gauge field $V = 1$. The correction $V_1$ due to exchange of one gauge boson is
\begin{eqnarray*}
V_1 & & \sim \int_{\vec q, \Omega} \left(\vec v_F.\vec K_b - \left(\vec v_F. \hat{q}\right)\left(\vec K_b. \hat{q}\right)\right) \\
&& \times {\cal D}(\vec q, \Omega){\cal G}_f(\vec K_f - \vec q, \omega_f - \Omega) \\
&& \times {\cal G}_b(\vec K_b + \vec q, \omega_b + \Omega)
\end{eqnarray*}
where ${\cal D}$ is the gauge propagator, and ${\cal G}_{b,f}$ are the spinon and boson propagators. $\vec v_F$ is a vector normal to the Fermi surface and of magnitude $v_{F0}$.
For small $\vec K_b$ we will show that $V_1 \sim \vec K_b. \vec A$ with $\vec A$ finite as all the other small external momenta and frequencies go to zero. To that end let us examine the integral that determines $\vec A$ from the equation for $V_1$ above. We choose $\vec K_f = K_f \hat{x}, \omega_f = 0$.
Then
\begin{eqnarray*}
 \vec A & & \sim \int_{\vec q, \Omega} \left(\vec v_F - \left(\vec v_F. \hat{q}\right) \hat{q}\right) \\
&& \times {\cal D}(\vec q, \Omega){\cal G}_f(\vec K_f - \vec q, - \Omega) \\
&& \times {\cal G}_b( + \vec q,  \Omega)
\end{eqnarray*}
As in the calculation of the self energy above due to the presence of the spinon Fermi surface the important $q$ region has $q_x \sim q_y^2$ so that $|q_x| \ll |q_y|$.  Then $\hat{q}$ is nearly perpendicular to $\hat{v_F}$ so that $\left(\vec v_F - \left(\vec v_F. \hat{q}\right) \hat{q}\right)$ can be replaced by $\vec v_F$ so that $\vec A$ points along $\hat{x}$. Further we may replace $|q|$ in the boson and gauge propagators by $|q_y|$. Then we may do the $q_x$ integral to get
\begin{equation}
A_x \sim i\int_{\Omega, q_y} \left(\frac{sgn(-\Omega)}{\frac{k_0 |\Omega|}{|q_y|}+\sigma_0 c |q_y|}\right)
\frac{1}{\left((\omega_b + \Omega)^2 + c^2 q_y^2 \right)^{\frac{2-\eta}{2}}}
\end{equation}
Clearly $A_x(\omega_b = 0) = 0$. For small $\omega_b$ the integrals can be evaluated to show that
$A_x \sim isgn(\omega_b)|\omega_b|^\eta$. Thus the vertex correction $V_1 \sim \vec k_b. \vec A$ goes to zero for small boson momenta and can be ignored in the scaling limit.

\end{appendix}

\end{document}